\documentstyle[11pt]{article}                                          
\begin{document}

\begin{center}
{\Large\bf Klauder's coherent states for the radial Coulomb problem\\ in a 
uniformly curved space and their flat-space limits} \\[8mm]

Myo Thaik and Akira Inomata\footnote{E-mail address: inomata@albany.edu} 
\\[2mm]

Department of Physics, State University of New York at Albany, 
Albany, New York 12222
\end{center}
                         
{\small First a set of coherent states \'a la Klauder is formally 
constructed for the Coulomb problem in a curved space of constant 
positive curvature. Then the flat-space limit is taken to reduce 
the set for the radial Coulomb problem to a set of hydrogen atom 
coherent states corresponding to both the discrete and the continuous 
portions of the spectrum for a fixed $\ell $ sector.}  

\section{Introduction}

In \cite{Klau}, Klauder proposed a set of coherent states in relation 
with the bound state portion of the hydrogen atom, generalizing the 
harmonic oscillator coherent states so as to preserve the following 
three properties; they are (i) continuous in their parameters, (ii) 
admit a resolution of unity, and (iii) are temporally stable (i.e., 
evolve among themselves in time). In fact there are a number of ways to 
generalize the harmonic oscillator coherent states \cite{KS, Per}. Most 
generalizations, notably Perelomov's, are based on group structures. 
Klauder stipulates coherent states without resort to a group. 
 
The proposed coherent states with an energy 
spectrum $E_{n}=\omega e_{n}$~$(n=0, 1, 2, ...; e_{0}=0)$ are labeled by 
two real parameters $s~ (0 \leq s < \infty )$ and $\gamma ~(- \infty < 
\gamma < \infty )$ as 
\begin{equation} 
|s, \gamma \rangle = 
M(s^{2})\,\sum_{n=0}^{\infty }\frac{s^{n}e^{-i\gamma e_{n}}}{\sqrt{\rho 
_{n}}}\,|n \rangle   \label{Kcs} 
\end{equation} 
where $|n\rangle $ is the eigenstate belonging to $E_{n}$ and 
$\rho _{n}$ is the $n$th moment of a probability distribution function 
$\rho (u) >0$,
\begin{equation}
\rho _{n}=\int_{0}^{\bar{u}} u^{n}\,\rho (u)\,du. \label{rho} 
\end{equation}  
For the harmonic oscillator, $\rho (u)=e^{-u}$ leads to the desirable 
result $\rho _{n}=n!$. However, in general, the coherent states 
(\ref{Kcs}) as proposed by Klauder  have ambiguity in the choice of 
$\rho (u)$. The normalization factor $M(s^{2})$ is determined so as to 
satisfy $\langle s, \gamma |s, \gamma \rangle =1$; namely, 
\begin{equation} 
M(s^{2})^{-2}=\sum_{n=0}^{\infty }\frac{s^{2n}}{\rho _{n}}.
\end{equation} 

With the Hamiltonian $\hat{H}$ such that $\hat{H}|n\rangle = \omega 
e_{n}|n\rangle $, it is apparent that  
\begin{equation} 
e^{-i\hat{H}t}|s,\gamma \rangle = |s, \gamma + \omega t \rangle 
\end{equation} 
which is taken in \cite{Klau} as the exhibition of temporal stability of 
the coherent states. The states satisfy the resolution of unity, 
\begin{equation} 
\int \,d\mu (s, \gamma ) \,|s,\gamma \rangle \langle s,\gamma | 
= \hat{1}_{dis}
\end{equation} 
with a measure $\mu (s, \gamma )$ defined by 
\begin{equation} 
\int \,d\mu (s,\gamma )\,f(s,\gamma ) = \lim_{\Gamma \rightarrow \infty 
}\frac{1}{2\Gamma }\int_{0}^{\infty } k(s^{2})\,ds^{2}\, \int_{-\Gamma }^{\Gamma 
}\, d\gamma \, f(s,\gamma )  \label{meas}
\end{equation} 
provided that 
\begin{equation} \lim_{\Gamma \rightarrow \infty } \frac{1}{2\Gamma 
}\int_{-\Gamma }^{\Gamma }\,d\gamma \,e^{i\gamma (e_{n}-e_{n'})} = 
\delta _{n,n'}, 
\end{equation}                                       
that is, that all $e_{n}$ are distinct (no degeneracies). In 
(\ref{meas}),
\begin{equation}
k(s^{2})=\rho (s^{2})/M(s^{2})^{2},
\end{equation}
which remains unspecified until the form of $\rho (s^{2})$ in 
(\ref{rho}) is given. Gazeau and Klauder \cite{GK}, letting $s^{2}=J$, 
imposed an additional condition, called the action identity \cite{GK}, 
\begin{equation} 
\langle J, \gamma |\hat{H}|J, \gamma \rangle 
=\omega J,  \label{act} 
\end{equation} 
which leads $\rho _{n}$ to the form,
\begin{equation} 
\rho _{n}=\prod_{j=1}^{n} e_{j}, ~~~~~~~~\rho _{0}=1. 
\end{equation} 
This condition suggests one to interpret the parameter $J$ as the classical 
action variable conjugate to the angle variable $\gamma $. A remark will 
be made in Sec. 4 concerning a possible use of $J$ for the semiclassical 
quantization condition. 

Gazeau and Klauder \cite{GK} also proposed coherent states for continuum 
dynamics. For a Hamiltonian with a non-degenerate continuous spectrum $0 
< \omega \varepsilon  < \omega \bar{\varepsilon }$, the proposed 
coherent states take the form, 
\begin{equation}
|s, \gamma \rangle = M(s^{2})\,\int_{0}^{\bar{\varepsilon 
}}\frac{s^{\varepsilon }e^{-i\gamma \varepsilon }}{\sqrt{\rho 
(\varepsilon )}}|\varepsilon  \rangle d\varepsilon ,   \label{GKcont}                            
\end{equation} 
where 
\begin{equation}
M(s^{2})^{-2} = \int_{0}^{\bar{\varepsilon }}\frac{s^{2\varepsilon } 
}{\rho (\varepsilon )} d\varepsilon 
\end{equation}
to meet $\langle s, \gamma |s, \gamma \rangle =1$ for $0 \leq s < 
\bar{s}$. The function $\rho (\varepsilon )$ in (\ref{GKcont}) is determined 
with an appropriate non-negative weighting function $\sigma (s)\geq 0 $ as 
\begin{equation}
\rho (\varepsilon )=\int_{0}^{\bar{\varepsilon }} s^{2\varepsilon } \sigma (s)\,ds .  \label{sigma}
\end{equation}
These coherent states for a continuous spectrum evolve in time among 
themselves. With $d\mu (s,\gamma )=(1/2\pi )M(s)^{-2}\sigma 
(s)\,ds\,d\gamma $, the resolution of unity, 
\begin{equation}
\int d\mu (s,\gamma )|s, \gamma \rangle \langle s, \gamma |= \hat{1}_{cont},
\end{equation}
is fulfilled. In \cite{GK}, the resolution of unity is set up 
independently for the discrete and the continuous case.

In the present paper, we first consider within the Gazeau-Klauder 
framework a set of coherent states for the radial Coulomb problem in a 
curved space of constant positive curvature. Then, taking the flat-space 
limit, we obtain a set of coherent states for the continuous part as well 
as the discrete portion of the hydrogen spectrum in a unified manner. 
Although the spectrum of the Coulomb system in the curved space is wholly 
discrete, the set of coherent states we derive in the flat-space limit 
consists of the discrete and continuous portions. In particular, for the 
$S$-waves, the set coincides with the one constructed by Gazeau and 
Klauder for the hydrogen-like spectrum \cite{GK} if the continuous 
portion is ignored. 

\section{The Coulomb problem in a uniformly curved space}

Schr\"odinger \cite{Sch} was the first to find quantum mechanical 
solutions for the Coulomb problem in a curved space of constant positive 
curvature. He considered this problem as an example that can be solved 
by the factorization procedure but is not tractable by other methods 
(see also \cite{Inf}). Soon after, however, Stevenson \cite{Stev} 
succeeded in obtaining the solutions by a conventional manner. Indeed 
there are various ways to approach the problem. It may be worth 
mentioning that the same problem has been solved by the dynamical group 
approach \cite{BW} and by the path integral approach \cite{BIJ}. It is 
interesting that the system has only a discrete spectrum unlike the 
usual hydrogen atom in flat space and may be seen as a compactified 
version of the usual Coulomb problem. It is natural to expect that the 
discrete spectrum of the system will generate the entire spectrum of the 
hydrogen atom including both the continuous and the discrete portions 
when the curvature of the space diminishes. We shall explore this 
limiting property later. First we wish to construct the coherent states 
\'a la Klauder for the radial part of the Coulomb system. 

We assume that space is uniformly curved with a positive curvature 
$K=1/R^{2} >0$. Then the curved space may be realized as a 
three-dimensional sphere $(S^{3})$ of radius $R$ imbedded in a 
four-dimensional Euclidean space. The line element $ds$ of the space is 
given in polar coordinates by 
\begin{equation}
ds^{2} = \frac{dr^{2}}{1 - r^{2}/R^{2}} + r^{2}(d\theta ^{2} + \sin 
^{2}\theta \,d\phi ^{2}). 
\end{equation}  
Or, with $\sin \chi =r/R ~(\chi \in [0, \pi ])$, it can be put in the form,
\begin{equation}
ds^{2} = R^{2}d\chi ^{2} + R^{2}\sin^{2}\chi (d\theta ^{2} + \sin ^{2}\theta \,d\phi 
^{2}). \label{ds}
\end{equation}
The Coulomb potential on the sphere \cite{Sch} is 
\begin{equation}
V(\chi )=-\frac{Ze^{2}}{R}\cot \chi  \label{pot}
\end{equation}
which satisfies the harmonic condition \cite{Inf},
\[
\frac{d}{d\chi }\left(\sin ^{2}\chi \frac{dV}{d\chi }\right)=0. 
\]

The Hamiltonian operator for this Coulomb system with a unit mass 
is given by
\begin{equation}
\hat{H}=-\frac{1}{2}\hat{\Delta } - \frac{Ze^{2}}{R}\cot \chi, 
\end{equation}
where $\hbar=1$ and $\hat{\Delta }$ is the Laplace-Beltrami operator of 
$SO(4)$.  
The corresponding Schr\"odinger equation may be expressed as 
\begin{equation}
\left[\frac{1}{\sin ^{2}\chi }\frac{\partial}{\partial \chi }\left(\sin 
^{2}\chi 
\frac{\partial}{\partial \chi }\right) - \frac{\hat{L}^{2}}{\sin^{2} 
\chi} + 2\sqrt{2 \omega } R\, \cot \chi + 2 R^{2}E \right]
\psi (\chi , \theta , \phi )=0 
\end{equation}
where $\omega = Z^{2}e^{4}/2$, and $\hat{L}^{2}$ is the Casimir 
invariant of $SO(3)$,
\begin{equation}
\hat{L}^{2}=-\left[\frac{1}{\sin \theta }\frac{\partial}{\partial \theta }
\left(\sin \theta \frac{\partial}{\partial \theta }\right) + 
\frac{1}{\sin ^{2}\theta }\frac{\partial^{2}}{\partial \phi 
^{2}}\right].
\end{equation}
It is obvious that the $SO(3)$ portion can be separated by letting 
the wave function as the product of the radial function and the 
spherical harmonics, $ \psi (\chi , \theta , \phi ) \sim w_{\ell}(\chi 
)\,Y^{m}_{\ell}(\theta , \phi) $. 
The radial function $w_{\ell }(\chi ) $ obeys
\begin{equation}
(\hat{H}_{\ell } - E)\,w(\chi )=0
\end{equation}
with the radial Hamiltonian,
\begin{equation}
\hat{H}_{\ell }=-\frac{1}{2R^{2}}
\left[\frac{\partial^{2}}{\partial \chi ^{2}} + 2\cot \chi 
\frac{\partial }{\partial \chi } - \frac{\ell (\ell +1)}{\sin^{2} 
\chi} + 2\sqrt{2\omega }R\, \cot \chi \right].
\end{equation}
From this follows the degenerate energy spectrum \cite{Sch, Inf, Stev}
\begin{equation}
E_{N}=\frac{N^{2}-1}{2 R^{2}} - 
\frac{\omega }{N^{2}}, ~~~~~~(N=1, 2, 3, ...),  \label{spec1}
\end{equation} 
and the corresponding eigenfunctions \cite{Inf, Stev, BIJ}
\begin{equation}
w_{N, \ell }(\chi) \sim \sin^{\ell }\chi \,e^{-i\chi (N-\ell -1
+ i\lambda _{n})}\,\,_{2}F_{1}(\ell - N +1, \ell +1 -i\lambda 
_{n}; 2\ell +2; 1-e^{2i\chi }). \label{w1}
\end{equation}
In the above, $_{2}F_{1}(\alpha , \beta ; \gamma ; z)$ is Gauss's hypergeometric 
function, and 
\begin{equation}
\lambda _{n} = - \frac{R}{a(n+\ell +1)}= - \frac{R}{aN},
\end{equation} 
with $a=(2\omega )^{-1/2}=(Ze^{2})^{-1}$.

At this point we note that the present polar coordinate realization of 
the system, even being in a curved space, is not degeneracy-free. Since 
an analog of the Runge-Lentz vector exists and commutes with the 
Hamiltonian, the accidental degeneracy persists \cite{Higgs, Leemon}. 
Therefore, we consider only the coherent states associated with the 
radial wave functions. Fixing $\ell $ we label the wave functions by the 
radial quantum number $n=0,1,2,...$ rather than the principal quantum 
number $N=n+\ell +1=1,2,3,...$. 

With the radial quantum number $n$, the energy spectrum (\ref{spec1}) 
and the wave functions (\ref{w1}) may be given, respectively, by 
\begin{equation}
E_{n}=\frac{(n+\ell )(n+\ell +2)}{2R^{2}} - 
\frac{\omega }{(n+\ell +1)^{2}}, ~~~~~~(n=0, 1, 2, ...),  \label{spec2}
\end{equation} 
and 
\begin{equation}
w_{n, \ell }(\chi)=C_{n, \ell } \sin^{\ell }\chi \,e^{-i\chi (n
+ i\lambda _{n})}\,\,_{2}F_{1}(-n, \ell +1 -i\lambda 
_{n}; 2\ell +2; 1-e^{2i\chi }) \label{w2}
\end{equation}
with the normalization factor \cite{BIJ}
\begin{equation}
C_{n, \ell }=e^{i\pi (2n +\ell +1 )/2}\frac{2^{\ell +1}}{\Gamma (2\ell 
+2)}\left[\frac{i\{(n+\ell +1)^{2} + \lambda _{n}^{2}\} \Gamma (\ell 
+ 1 + i\lambda _{n})\Gamma (n+2\ell +2) }{R^{3}\,\kappa _{n}\,\Gamma 
(i\lambda _{n} - \ell )\Gamma (n +1 )}\right]^{1/2}
\end{equation}
where  
\begin{equation}
\kappa _{n} = \mbox{min }\{|n+\ell +1|, |\lambda _{n}|\}.
\end{equation}

Now we write down the coherent states \'a la Klauder for the radial 
Coulomb problem on the sphere; namely, 
\begin{equation}
|s, \gamma \rangle = {\cal M}(s^{2})\,\sum_{n=0}^{\infty 
}\frac{s^{n}e^{-i\gamma [n]_{R}}}{\sqrt{[n]_{R}!}}\,|n \rangle .
\label{Ccs}
\end{equation}
Here we note that $|n\rangle $ is the $n$th energy eigenvector so that 
$w_{n,\ell}(\chi )=\langle \chi |n\rangle $. Also we have 
defined a generalized number $[n]$ by 
\begin{equation}
[n]=(E_{n}-E_{0})/\omega ~~~~~~~(n=0,1,2,...).
\end{equation}
and $[0]!=1$. The subscript $R$ stands for a finite radius $R$ of curvature. 
For (\ref{spec2}), we have   
\begin{equation}
[n]_{R}!=\prod_{m=1}^{n}[m]_{R}=\prod_{m=1}^{n}\left[ 
\frac{m(m+2\ell +2)}{(m+\ell +1)^{2}(\ell +1)^{2}}\left(1 + 
\frac{(m+\ell +1)^{2}(\ell +1)^{2}}{2 \omega R^{2}}\right) \right]
\end{equation}
and 
\begin{equation}
{\cal M}(s^{2})^{-2}= \sum_{n=0}^{\infty }\frac{s^{2n}}{[n]_{R}!}.
\end{equation}
Since $[n]_{R}!$ and hence ${\cal M}(s^{2})$ cannot be given in 
closed form, the coherent states (\ref{Ccs}) so constructed 
for the radial Coulomb problem in curved space are not particularly 
interesting until their flat space limits are taken. Nevertheless, it 
is obvious that the set of coherent states given above possess all the 
properties (i)-(iii) of Klauder's coherent states \cite{Klau} plus the 
action identity of Gazeau and Klauder \cite{GK}. 

\section{The coherent states for the radial Coulomb problem in flat space} 

Next we consider the flat-space limits where the radius of curvature $R$ 
tends to infinity. We conjecture that the Coulomb problem on the sphere 
with $Z=1$ goes over to the hydrogen atom problem (with $m_{e}=1$) in 
the flat space limit. By doing so, we expect that the discrete energy 
spectrum of the Coulomb problem on the sphere will correspond to both 
the discrete and the continuous parts of the hydrogen atom spectrum in 
flat space \cite{BIJ}. 

Before taking the limit, we introduce the critical number $n_{c}$ for 
which the energy becomes zero, that is, $E_{n_{c}}=0$, or
\begin{equation}
(n_{c}+\ell )(n_{c} + \ell + 2)(n_{c} + \ell +1)^{2}=2\omega R^{2}.
\label{nR}
\end{equation}
Then we separate the spectrum (\ref{spec2}) into two parts: (a) $E_{n} < 
0 $  and (b) $ E_{n} \geq 0$, and consider their limiting cases 
separately.\\   

\noindent \underline{Case (a)} ~$E < 0$ ($n < n_{c}$):~It is apparent 
from (\ref{nR}) that  
as $R$ approaches infinity $n_{c} $ goes to infinity as fast as 
$\sqrt{R}$. Accordingly the first term of the energy 
spectrum (\ref{spec2}) for $n < n_{c}$ tends to zero as 
\begin{equation}
\left|E_{n} + \frac{\omega }{(n+\ell +1)^{2}}\right| = 
\frac{(n + \ell )(n+\ell + 2)}{2mR^{2}} <  \frac{(n_{c}+\ell 
)(n_{c}+\ell +2)}{2mR^{2}} \sim \frac{1}{R} \rightarrow 0.
\end{equation}
Since $n_{c} \rightarrow \infty $, the energy 
spectrum bounded above by zero takes the form,  
\begin{equation}
E_{n}= - \frac{\omega }{(n+\ell +1)^{2}} ~~~~(n=0,1,2,...) \label{dspec}
\end{equation}
which coincides, as is expected, with the discrete hydrogen atom 
spectrum.  

Now notice that Gauss's hypergeometric function is reduced to Kummer's 
confluent hypergeometric function by the limiting procedure,
\[
\displaystyle \lim_{\beta \rightarrow \infty } \, _{2}F_{1}(\alpha , \beta 
; \gamma ;z/\beta )= \,_{1}F_{1}(\alpha ; \gamma  ;z)
\]
and that for large $|z|$
\[
\frac{\Gamma (z+\ell +1)}{\Gamma (z-l)} \sim z^{2\ell +1}.
\]
Recalling also that $\lambda _{n} = -R/\{a(n+\ell +1)\}$, we obtain the 
following limiting values for $E < 0 $, 
\begin{equation}
\begin{array}{ll}
\displaystyle \lim_{R\rightarrow \infty }\,_{2}F_{1}(-n-1, \ell +1-
i\lambda _{n}; 2\ell +2; 1-e^{2i\chi })=\displaystyle \,_{1}F_{1}(-n -
1; 2\ell + 2;2r/a(n+\ell +1)) 
& ~\\
& ~\\
\displaystyle \lim_{R\rightarrow \infty }\,\exp[-i\chi  (n +1 +i\lambda _{n})]
=\exp[-r/a(n+\ell +1)] & ~\\
&~\\
\displaystyle \lim_{R\rightarrow \infty 
}\,\sin^{\ell }\chi \left[\frac{i\{(n+\ell +1)^{2}+\lambda 
_{n}^{2}\}\,\Gamma (\ell +1 +i\lambda _{n})}{R^{3}(n+\ell +1) \,\Gamma 
(i\lambda _{n}-\ell) }\right]^{1/2} = 
\left[\frac{ir}{a(n+\ell +1)}\right]^{\ell 
}\left[\frac{1}{a^{3}(n+\ell +1)^{4}}\right]^{1/2}. & ~ \label{lim1}
\end{array}
\end{equation}
Thus, in the flat-space limit, the radial function (\ref{w1}) takes the 
form, 
\begin{equation}
u_{n,\ell }(r)=C_{n}
\left[\frac{2r}{a(n+\ell +1)}\right]^{\ell }e^{-r/a(n+\ell +1)} 
\,_{1}F_{1}(-n-1; 2\ell +2; 2r/a(n+\ell +1)) \label{u-func}
\end{equation}
with the normalization constant,
\begin{equation}
C_{n}=
\frac{1}{(2\ell +1)!}\left[\left(\frac{2}{a(n+\ell +1)}\right)^{3}
\frac{(n+2\ell +1)!}{2(n+\ell +1)(n+1)!}\right]^{1/2} .
\end{equation}         
This result is in fact the normalized hydrogen atom radial wave 
function in units where $\hbar=1$. See, e.g., \cite{Merz}.\\ 
 
\noindent \underline{Case (b)} ~$E \geq 0$ ($n \geq n_{c}$):~
For large $R$, we approximate $\Delta n /R$ for $n > n_{c}$ by $dk$ with 
a continuous parameter $k > 0$, so that by integration 
\begin{equation}
n - n_{c} =kR.
\end{equation} 
In the limit $R \rightarrow \infty $, the energy spectrum behaves as 
\begin{equation}
E_{n}=\frac{(kR + n_{c} + \ell)(kR + n_{c} + \ell +2)}{2R^{2}} - \frac{\omega 
}{(kR + n_{c} + \ell +1)^{2}} \rightarrow \frac{k^{2}}{2}.
\label{cspec}
\end{equation}
As a result, the discrete spectrum (\ref{spec2}) for $E \geq 0$ 
turns into a continuous spectrum, 
\begin{equation}
E(k)=\frac{k^{2}}{2}  ~~~~(0\leq k ).
\end{equation}
In this continuous case, for large $R$, we must replace $\lambda 
_{n} $ by $-1/ak$. In a way similar to evaluating the discrete limits
(\ref{lim1}), we calculate the continuous limiting values,  
\begin{equation}
\begin{array}{ll}
\displaystyle \lim_{R\rightarrow \infty }\,_{2}F_{1}(\ell -n, \ell +1-
i\lambda _{n}; 2\ell +2; 1-e^{2i\chi })=\displaystyle  _{1}F_{1}(\ell 
+ 1 + i/ak; 2\ell + 2; 2ikr) 
& ~\\
& ~\\
\displaystyle \lim_{R\rightarrow \infty }\,\exp[-i\chi  (n-\ell +
i\lambda _{n})]=\exp[-ikr] & ~\\
& ~\\
\displaystyle \lim_{R\rightarrow \infty 
}\,\sin ^{\ell }\chi \,\left[\frac{i}{R^{3}}\frac{((n+1)^{2}+\lambda 
_{n}^{2})\,\Gamma (n + \ell + 2)}{|\lambda _{n}| \,\Gamma 
(n -\ell + 1)}\right]^{1/2}=2^{-\ell }e^{i\pi 
/4}\sqrt{a}\,k^{2}\,(2kr)^{\ell }. 
\end{array}
\end{equation}
Using these results, we arrive at the limiting wave function belonging 
to the continuous spectrum for $E \geq 0$, 
\begin{equation}
v_{k, \ell }(r)=\left(\frac{2a}{\pi }\right)^{1/2}\frac{k^{2}(2kr)^{\ell 
}}{(2\ell +1)!}|\Gamma (\ell +1 -i/ak)|\,\sinh^{1/2}(\pi /ak)\,e^{ikr}
_{1}F_{1}(\ell +1 +i/ak; 2\ell +2; 2ikr).
\end{equation}
In this manner, we see that the discrete dynamics of the Coulomb system in 
curved space leads to the discrete and continuous  
regimes of the hydrogen atom in flat space. 

Now we consider the flat-space limit of the coherent states (\ref{Ccs})
for fixed $\ell $:
\begin{equation}
|s, \gamma \rangle = \lim_{R\rightarrow \infty }\,
{\cal M}(s^{2})\left[\sum_{E < 0 } + \sum_{E \geq 0 } \right ] 
\,\frac{s^{n}\,e^{-i\gamma [n]}}{\sqrt{[n]_{R}!}}|n \rangle .   
\end{equation}

Corresponding to the limiting discrete spectrum (\ref{dspec}) for $E < 
0$, we write
\begin{equation}
\lim_{R\rightarrow \infty }[n]_{R}=[n]
\end{equation} 
and have
\begin{equation}
[n]!
=\prod_{m=1}^{n}\frac{m(m+2\ell +2)}{(\ell +1)^{2}(m+\ell +1)^{2}}
=\frac{n!}{(\ell +1)^{2n}}\frac{(2\ell +3)_{n}}{[(\ell 
+2)_{n}]^{2}}, \label{n}
\end{equation}
where $(z)_{n}$ is the Pochhammer symbol,
\begin{equation}
(z)_{n}=\Gamma (z+n)/\Gamma (z), ~~~~~~~(z)_{0}=1.
\end{equation}
In particular, for the $S$-wave case ($\ell =0$), this reduces to 
the result given for the Coulomb-like spectrum in \cite{GK}:
\begin{equation}
\rho _{n}=\frac{n+2}{2(n+1)}.
\end{equation}

With (\ref{n}) the discrete portion of the coherent states becomes 
\begin{equation}
|s, \gamma \rangle _{disc}= {\cal N}(s^{2})\sum_{n=0}^{\infty } 
\,\frac{s^{n}\,e^{-i\gamma [n]}}{\sqrt{[n]!}}
|n \rangle _{disc}, \label{CSdis}          
\end{equation}
where 
\begin{equation}
{\cal N}(s^{2}) = \lim_{R \rightarrow \infty } {\cal M}(s^{2})
\end{equation}
which will be evaluated shortly. The discrete eigenstates $|n\rangle$ and 
the radial wave functions (\ref{u-func}) with a fixed $\ell $ are 
related by $u_{n,\ell }(r)=\langle r|n\rangle $. Naturally the result 
(\ref{CSdis}) for the discrete portion coincides with Klauder's coherent 
state (\ref{Kcs}) except for the normalization factor. 

For the continuous spectrum (\ref{cspec}) for $E\geq 0$, adopting the
weighting function $\sigma (s)=e^{-s}$ in (\ref{sigma}), we take the 
limiting value, 
\begin{equation}
\lim _{R \rightarrow \infty }[n]_{R}! = \rho (\varepsilon )
=\int_{0}^{\infty } s^{\varepsilon }\,e^{-s}\,ds = 
\Gamma \left[\varepsilon +1 \right] =\varepsilon !,  \label{Gamma}
\end{equation} 
where $\varepsilon = E(k)/\omega = k^{2}/(2\omega ).$ Note that writing 
$\Gamma (\varepsilon +1)$ formally as $\varepsilon !$ in (\ref{Gamma}) 
is to stress that it is a natural continuum limit of $[n]!$. 

Then the continuous portion of the coherent states may be constructed 
in the form, 
\begin{equation} 
|s, \gamma  \rangle _{cont} = {\cal N}(s^{2})
\int_{0}^{\infty }\,d\varepsilon \, \frac{s^{\varepsilon }\,e^{-i \gamma \varepsilon  
}}{\sqrt{\Gamma (\varepsilon +1)}} |\varepsilon \rangle _{cont},  
\end{equation} 
expanded with the states $|\varepsilon \rangle $ satisfying 
\begin{equation}
\hat{H}_{\ell }|\varepsilon \rangle = \omega \varepsilon |\varepsilon 
\rangle , ~~~~~~
\langle \varepsilon |\varepsilon '\rangle =\delta (\varepsilon -\varepsilon ').
\end{equation}
The normalization factor ${\cal N}(s^{2})$ common to the discrete and 
continuous portions is given by 
\begin{equation}
{\cal N}(s^{2})^{-2}=\sum_{n=0}^{\infty }\frac{s^{2n}}{[n]!} + 
\int_{0}^{\infty }d\varepsilon \,\frac{s^{2\varepsilon }}{\varepsilon 
!},
\end{equation}
which can be cast into the form,
\begin{equation}
{\cal N}(s^{2})^{-2}= \,_{2}F_{1}\left(\ell +2, \ell +2; 2\ell +3; \,(\ell 
+1)^{2}s^{2}\right) + \nu (s^{2}),
\end{equation}
where $\nu (s)$ is the $\nu $-function \cite{GR} defined by 
\begin{equation}
\nu (x) = \int_{0}^{\infty } \frac{x^{t}}{\Gamma (t+1)}\,dt. 
\end{equation}

Consequently, as the flat-space limit of the coherent states (\ref{Ccs}) 
for the Coulomb problem in curved space, we obtain the coherent states 
for the radial Coulomb problem consisting of the discrete and continuous 
portions:
\begin{equation}
|s, \gamma \rangle = {\cal N}(s^{2}) \left[\sum_{n=0}^{\infty } 
\,\frac{s^{n}\,e^{-i\gamma [n]}}{\sqrt{[n]!}}
|n \rangle _{disc} + 
\int_{0}^{\infty }\,d\varepsilon \, \frac{s^{\varepsilon }e^{-i \gamma \varepsilon 
}}{\sqrt{\varepsilon !}}
|\varepsilon  \rangle _{cont} \right].   \label{2cs}
\end{equation}

In particular, for the $S$-wave sector ($\ell =0$), we have
\begin{equation}
|s, \gamma \rangle _{\ell =0}= {\cal N}_{0}(s^{2}) \left[\sum_{n=0}^{\infty } 
\,\frac{s^{n}\,e^{-i\gamma n(n+2)/(n+1)^{2}}}{\sqrt{(n+2)/(2n+2)}}
|n, \ell =0 \rangle _{disc} + 
\int_{0}^{\infty }\,d\varepsilon \, \frac{s^{\varepsilon }e^{-i \gamma \varepsilon 
}}{\sqrt{\varepsilon !}}
|\varepsilon , \ell =0 \rangle _{cont} \right],   \label{3cs}
\end{equation}
with 
\begin{equation}
{\cal N}_{0}(s^{2})^{-2}=\frac{2}{s^{2}}\left[\frac{s^{2}}{1-s^{2}} + \ln (1 
- s^{2})\right] + \nu (s^{2}).
\end{equation}
The discrete portion of this $S$-wave limit coincides with the result 
given by Gazeau and Klauder \cite{GK}.  

The set of coherent states just obtained in the flat space limit possesses 
all the properties (i)-(iii) posed by Klauder \cite{Klau}. Naturally
the resolution of unity is extended to include both the discrete and 
continuous states: 
\begin{equation}
\int  |s, \gamma \rangle \langle s, \gamma |\, d\mu (s, \gamma ) = {\bf 
1}, 
\end{equation} 
with the same measure as that of (\ref{meas}). This relation is not 
valid when the continuous states are ignored. In addition, the action 
identity of Gazeau and Klauder \cite{GK} is satisfied: 
\begin{equation}
\langle s, \gamma |\hat{H}_{\ell }-E_{0}|s, \gamma \rangle
=\omega J   \label{oJ}
\end{equation}
with the identification $J=s^{2}$.

\section{Concluding Remarks}

We have obtained both the discrete and continuous portions of the 
coherent states for the radial Coulomb problem in a unified manner.  
We emphasize that the coherent states for the discrete spectrum in a 
uniformly curved space are reducible in the flat space limit to those 
for the continuous spectrum plus those for the discrete spectrum. In the 
$S$-wave limit $(\ell =0)$, if the continuous part is ignored, our 
result (\ref{2cs}) coincides with that of Gazeau and Klauder \cite{GK} 
for the Coulomb-like discrete spectrum. Although the classical 
action-angle interpretation of $J=s^{2}$ and $\gamma $ offered by Gazeau 
and Klauder is natural in (\ref{oJ}) for the harmonic oscillator case, 
it is questionable that such an interpretation is appropriate for the 
Coulomb case. If $\langle J, \gamma |\hat{H}|J, \gamma \rangle $ 
corresponds to the classical energy $E_{cl}$, then $J=E_{cl}/\omega $ is 
the adiabatic invariant for the oscillator (after an appropriate 
adjustment of its dimension). The Bohr-Ishiwara-Sommerfeld-Wilson 
quantization $J \rightarrow n$ applied to the harmonic oscillator leads 
to $E_{cl}\rightarrow n\omega $, suggesting that the eigenvalues of 
$\hat{H}$ are $n\omega$. However, $J$ does not seem to be an adiabatic 
invariant for any other systems in a strict sense. The semiclassical 
quantization $J \rightarrow n$ does not yield a correct spectrum for a 
system other than the harmonic oscillator. Nonetheless it is interesting 
to point out that the quantization condition, if modified as 
$J\rightarrow [n]$ (replacing the integer $n$ by the generalized number 
$[n]=e_{n}$), leads to $E_{cl} \rightarrow e_{n}\omega $, and is 
compatible with the action-angle interpretation.

\end{document}